\documentclass[preprint,aps]{article}

\begin{document}
\begin{titlepage}
\centerline{\large \bf Photoionization of I$^+$ ion}  
\vglue .1truein
\centerline{Zhifan Chen and Alfred Z. Msezane}
\centerline{\it Center for Theoretical Studies of Physical Systems,  and
Department of Physics}
\centerline{\it Clark Atlanta University, Atlanta, Georgia 30314, U. S. A.}
\centerline{}
\centerline{}
\centerline{\bf ABSTRACT}
Photoionization of the I$^+$ ion in the energy range of 
the 4d giant resonance has been studied using our recently developed
random-phase approximation with exchange (RPAE) method.   
Photoionization cross sections for the I$^+$ $4d-\epsilon f, \epsilon p$, $5s-\epsilon p$, $5p-\epsilon s,\epsilon d$ have been 
obtained for each term of the ground state.
Calculations include all the intra-shell and inter-shell coupling
among the $4d$, $5s$, and $5p$ subshells.
Our maximum cross section for the I$^+$ $4d$ giant resonance, 23.12 MB at 90.24 eV agrees excellently  
with the recent measurement,
23(3) at 90 eV. 

\centerline{}
\centerline{}

\end{titlepage}
\section{Introduction}
Recently, the photoionization of the ion Xe$^{q+}$(q=1-6) [1-9] and the atom I and its 
ions I$^{q+}$ (q=1-2) [10-15],
have become "hot topics". Firstly, this is because the 4d giant resonances of the above atom/ions
are the important photoionization processes. Secondly,
the newly developed RPAE method has made a full RPAE calculation possible for 
the relevant reactions. 
Thirdly, the photoionization cross sections for the ionic species 
are required for plasma modeling.  

The random-phase approximation with exchange (RPAE) method has been used to 
calculate the I$^+$ photoionization cross section before. However, 
the calculations [9, 12-13] applied a variety of approximations as the RPAE 
method was initially developed for the calculation of closed shell atoms only.  
The previous RPAE calculations referenced above cannot provide the cross sections for each 
individual term of the I$^+$ ground state either.

Recently, a RPAE method has been developed
by Chen and Msezane [8].   
The method has been successfully used to study the inner-shell electron transitions of an
atom with an outer open-shell. 
The calculation  
can include intra-shell
correlations, viz. an electron absorbs a photon then interacts and shares excitation energy
with other electrons in the same shell and 
the inter-shell correlation, where the electron absorbs a photon and interacts
and shares the excitation energy with the electrons of the neighboring
shells.  More resently a RPAE method,
which allows for the inclusion of both intra-shell and inter-shell correlations
has also been developed by Chen and Msezane for atoms(ions) 
with an inner open-shell [16]. These methods have greatly extended the 
scope of the RPAE method as their calculations allow the inclusion of all the inter-shell 
couplings among the Xe$^+$ $4d$, $5s$ and $5p$ subshells using the term-dependent
Hartree-Fock wave functions. 

In this paper we use our recently developed RPAE methods to study the photoionization of 
the I$^+$ ion. The difference between the current calculation and the previous RPAE calculations
[9, 12-13] is 
that the photoionization cross section for
each term of the
I$^+$ ground state, I$^+$ $4d^{10} 5s^2 5p^4$ $(^3P),\ (^1D), \ (^1S)$ has been calculated 
separately with the inclusion of all the inter-shell coupling among the I$^+$ $4d$, $5s$ and $5p$ 
subshells and without making any of the approximations used in the previous RPAE calculations.
The total cross section is evaluated as the weighted average sum of the cross sections 
for each term.
Our maximum for the I$^+$ $4d$ giant resonance shows an excellent agreement with the 
recent measurement [14].

\section{Theory}

The RPAE equation and corresconding dipole and Coulomb 
matrix elements for an atom with an outer open-shell 
is given by Eq. (1) and Appendix A  of Ref. [8]. 
The Coulomb interaction in Eq. (1) causes an electron to switch  
from $l_2$ to $l_1$. Similar terms will be added in the 
calculations for the electon switching from $l_1$ to $l_3$, or from $l_2$ to $l_3$ etc.  
of the state $|l_3^{n_3} [L_3 S_3] l_1^{n_1}[L_1 S_1] l_2^{n_2} [L_2 S_2] LS>$.   

The ground state of the I$^+$ atom has the configuration 
$4d^{10} 5s^2 5p^4 (^3P,\ ^1D,\ ^1S)$. Therefore our calculations have 
three parts, each calculating one term of the ground state.   

The following scattering processes are for  
the I$^+$ $4d^{10} 5s^2 5p^4 (^3P)$ term:

\bigskip
 
\begin{displaymath}
\ \ h\nu + 4d^{10} 5s^2 5p^4 (^3P)\rightarrow 4d^{9} 5s^2 5p^4 (^3P) (^2D,\ ^2F,\ ^4D,\ ^4F)
\epsilon f (^3P)
\end{displaymath}
\begin{displaymath}  
\ \ h\nu +4d^{10} 5s^2 5p^4 (^3P) \rightarrow  4d^{9} 5s^2 5p^4 (^3P)(^2P,\ ^2D,\ ^2F,\ ^4P,\ ^4D,\ ^4F)
\epsilon f (^3D)
\end{displaymath}   
\begin{displaymath}
\ \ h\nu +4d^{10} 5s^2 5p^4 (^3P) \rightarrow  4d^{9} 5s^2 5p^4 (^3P) (^2F,\ ^4F)
\epsilon f (^3S)
\end{displaymath}
\begin{displaymath}
\ \ h\nu + 4d^{10} 5s^2 5p^4 (^3P) \rightarrow 4d^{10} 5s^2 5p^3 (^2D) 
\epsilon s (^3D)
\end{displaymath}
\begin{displaymath}
\ \ h\nu + 4d^{10} 5s^2 5p^4 (^3P) \rightarrow 4d^{10} 5s^2 5p^3 (^2P) 
\epsilon s (^3P)
\end{displaymath}
\begin{displaymath}
\ \ h\nu + 4d^{10} 5s^2 5p^4 (^3P) \rightarrow 4d^{10} 5s^2 5p^3 (^4S)  
\epsilon s (^3S)
\end{displaymath}
\begin{displaymath}
\ \ h\nu + 4d^{10} 5s^2 5p^4 (^3P) \rightarrow 4d^{10} 5s^2 5p^3 (^2P,\ ^2D) 
\epsilon d (^3D)
\end{displaymath}
\begin{displaymath}
\ \ h\nu + 4d^{10} 5s^2 5p^4 (^3P) \rightarrow 4d^{10} 5s^2 5p^3 (^2P,\ ^2D)
\epsilon d (^3P)
\end{displaymath}
\begin{displaymath}
\ \ h\nu + 4d^{10} 5s^2 5p^4 (^3P) \rightarrow 4d^{10} 5s^2 5p^3 (^2D)
\epsilon d (^3S)
\end{displaymath}
\begin{displaymath}
\ \ h\nu + 4d^{10} 5s^2 5p^4 (^3P) \rightarrow 4d^{10} 5s^2 5p^3 (^4S) 
\epsilon d (^3D)
\end{displaymath}
\begin{displaymath}
\ \ h\nu + 4d^{10} 5s^2 5p^4 (^3P) \rightarrow 4d^{10} 5s 5p^4 (^3P) (^2P,\ ^4P)
\epsilon p (^3S)
\end{displaymath}
\begin{displaymath}
\ \ h\nu + 4d^{10} 5s^2 5p^4 (^3P) \rightarrow 4d^{10} 5s 5p^4 (^3P) (^2P,\ ^4P)
\epsilon p (^3P)
\end{displaymath}
\begin{displaymath}
\ \ h\nu + 4d^{10} 5s^2 5p^4 (^3P) \rightarrow 4d^{10} 5s 5p^4 (^3P) (^2P,\ ^4P)
\epsilon p (^3D)
\end{displaymath}
\begin{displaymath}
\ \ h\nu + 4d^{10} 5s^2 5p^4 (^3P)\rightarrow 4d^{9} 5s^2 5p^4 (^3P) (^2D,\ ^2F,\ ^2P,\ ^4D,\ ^4F,\ ^4P)
\epsilon p (^3D)
\end{displaymath}
\begin{displaymath}  
\ \ h\nu +4d^{10} 5s^2 5p^4 (^3P) \rightarrow  4d^{9} 5s^2 5p^4 (^3P) (^2P,\ ^4P,\ ^2D,\ ^4D)
\epsilon p (^3P)
\end{displaymath}   
\begin{displaymath}
\ \ h\nu +4d^{10} 5s^2 5p^4 (^3P) \rightarrow  4d^{9} 5s^2 5p^4 (^3P) (^2P,\ ^4P)
\epsilon p (^3S)
\end{displaymath}

The total of 13 $(^3P)$, 18 $(^3D)$, and 8 $(^3S)$ final states have been included in the calculation.

\ \ \ \ The following scattering processes are for the 
I$^+$ $4d^{10} 5s^2 5p^4 (^1D)$ term: 

\begin{displaymath}
\ \ h\nu + 4d^{10} 5s^2 5p^4 (^1D)\rightarrow 4d^{9} 5s^2 5p^4 (^1D) (^2D,\ ^2F,\ ^2G)
\epsilon f (^1P)
\end{displaymath}
\begin{displaymath}  
\ \ h\nu +4d^{10} 5s^2 5p^4 (^1D) \rightarrow  4d^{9} 5s^2 5p^4 (^1D) (^2P,\ ^2D,\ ^2F,\ ^2G)
\epsilon f (^1D)
\end{displaymath}   
\begin{displaymath}
\ \ h\nu +4d^{10} 5s^2 5p^4 (^1D) \rightarrow  4d^{9} 5s^2 5p^4 (^1D) (^2S,\ ^2P,\ ^2D,\ ^2F,\ ^2G)
\epsilon f (^1F)
\end{displaymath}
\begin{displaymath}
\ \ h\nu + 4d^{10} 5s^2 5p^4 (^1D) \rightarrow 4d^{10} 5s^2 5p^3 (^2P) 
\epsilon s (^1P)
\end{displaymath}
\begin{displaymath}
\ \ h\nu + 4d^{10} 5s^2 5p^4 (^1D) \rightarrow 4d^{10} 5s^2 5p^3 (^2D)  
\epsilon s (^1D)
\end{displaymath}
\begin{displaymath}
\ \ h\nu + 4d^{10} 5s^2 5p^4 (^1D) \rightarrow 4d^{10} 5s^2 5p^3 (^2P,\ ^2D) 
\epsilon d (^1P)
\end{displaymath}
\begin{displaymath}
\ \ h\nu + 4d^{10} 5s^2 5p^4 (^1D) \rightarrow 4d^{10} 5s^2 5p^3 (^2P,\ ^2D)
\epsilon d (^1D)
\end{displaymath}
\begin{displaymath}
\ \ h\nu + 4d^{10} 5s^2 5p^4 (^1D) \rightarrow 4d^{10} 5s^2 5p^3 (^2P,\ ^2D)
\epsilon d (^1F)
\end{displaymath}
\begin{displaymath}
\ \ h\nu + 4d^{10} 5s^2 5p^4 (^1D) \rightarrow 4d^{10} 5s 5p^4 (^1D) (^2D)
\epsilon p (^1D,\ ^1P,\ ^1F)
\end{displaymath}
\begin{displaymath}
\ \ h\nu + 4d^{10} 5s^2 5p^4 (^1D) \rightarrow 4d^{9} 5s^2 5p^4 (^1D) (^2D,\ ^2F,\ ^2P)
\epsilon p (^1D)
\end{displaymath}
\begin{displaymath}
\ \ h\nu + 4d^{10} 5s^2 5p^4 (^1D) \rightarrow 4d^{9} 5s^2 5p^4 (^1D) (^2P,\ ^2D,\ ^2S)
\epsilon p (^1P)
\end{displaymath}
\begin{displaymath}
\ \ h\nu + 4d^{10} 5s^2 5p^4 (^1D) \rightarrow 4d^{9} 5s^2 5p^4 (^1D) (^2D,\ ^2F,\ ^2G)
\epsilon p (^1F)
\end{displaymath}
The total of 10 $(^1P)$, 11 $(^1D)$, and 11 $(^1F)$ final states have been included in the calculation.

\bigskip
The following scattering processes are for the  
I$^+$ $4d^{10} 5s^2 5p^4 (^1S)$ term:
\begin{displaymath}
\ \ h\nu + 4d^{10} 5s^2 5p^4 (^1S)\rightarrow 4d^{9} 5s^2 5p^4 (^1S) (^2D)
\epsilon f (^1P)
\end{displaymath}
\begin{displaymath}
\ \ h\nu + 4d^{10} 5s^2 5p^4 (^1S) \rightarrow 4d^{10} 5s^2 5p^3 (^2P) 
\epsilon s (^1P)
\end{displaymath}
\begin{displaymath}
\ \ h\nu + 4d^{10} 5s^2 5p^4 (^1S) \rightarrow 4d^{10} 5s^2 5p^3 (^2P)
\epsilon d (^1P)
\end{displaymath}
\begin{displaymath}
\ \ h\nu + 4d^{10} 5s^2 5p^4 (^1S) \rightarrow 4d^{10} 5s 5p^4 (^1S) (^2S)
\epsilon p (^1P)
\end{displaymath}
\begin{displaymath}
\ \ h\nu + 4d^{10} 5s^2 5p^4 (^1S)\rightarrow 4d^{9} 5s^2 5p^4 (^1S) (^2D)
\epsilon p (^1P)
\end{displaymath}
A total of 5 $(^1P)$ final states have been included in the calculation.

The I$^+$ ground state and the core wave functions were obtained through the 
self-consistent 
Hartree-Fock (HF) calculation. Then the radial functions of the continuum electron were 
obtained by solving the linear HF equations without self-consistency using those core 
wave functions. Each channel includes three discrete and twenty 
continuum wave functions. Each radial part of the wave function was represented by 2000
points.  
  
After evaluating the dipole and Coulomb matrix elements, the RPAE equation was solved for the 
partial cross sections for the $^3P$, $^3D$, $^3S$, $^1P$, $^1D$, and $^1F$ final states.  
Then the photoionization cross sections for each term, $^3P$, $^1D$ and $^1S$ was obtained 
from the sum of the
partial cross sections. Finally the total photoionization cross section was evaluated as the
weighted average of each term. 

\section{Results}
The I$^+$ $4d -\epsilon f$ photoionization cross section is given in Fig.1. Dotted, dashed and 
dash-dot 
curves represent, respectively the photoionization cross sections from the  
term, $4d^{10} 5s^2 5p^4 (^3P)$, $(^1S,)$ and $(^1D)$. The three curves are very 
close to each other. 
Fig. 1. shows a broad region corresponding to the $4d -\epsilon f$ giant resonance.   
It is a
shape resonance which is  
mainly determined by the potential shape. Fig. 1. gives a maximum cross section of 20.64 MB 
at the photon energy 
of 90.24 eV.  
Since almost all of the $4d - \epsilon f$ transition will emit a electron through 
double Auger decay,
this data 
should be compared with the experimental data for the I$^{+3}$ ion.  

Fig. 2. shows 
the $4d -\epsilon p$ 
photoionization cross sections.   
Obiviously, the
$4d-\epsilon p$ cross sections are much smaller in comparison than those for 
the $4d-\epsilon f$ transition. 
The description of the curves is the same as that
in Fig. 1. At about 67 eV, which is near the 4d threshold the curve has a peak. Then, the cross 
section gradualy decreases as the energy increases.
Around 90 eV all three curves have a very small broadened peak. This is caused by the 
inter-shell coupling with the $4d-\epsilon f$ 
channel. After carefully analyzing the partial cross sections, we found that 
for the ground state term  
$4d^{10} 5s^2 5p^4 (^3P)$ the largest contribution to the cross section is  
from the transition to the final state $^3D$. However, for the term  
$4d^{10} 5s^2 5p^4 (^1D)$ the 
largest contributor to the cross section comes from the final state $^1F$.  

The photoionization cross sections for the $5p$ and $5s$ subshells  
are given, respectively in Fig. 3 and Fig. 4. Fig. 3 demonstrates 
the $5p- \epsilon d, \epsilon s$ 
photoionization results. At about 19.38 eV the cross section is 17.89 MB. 
Then the cross section 
decreases as the energy increases. At 62.65 eV the cross section has reduced to 0.44 MB. 
After entering 
the energy range of the 4d giant resonance the cross section increases and 
reaches a maximum.  
A similar situation applies to the 5s photoionization process. 
However, firstly the cross section increases as the energy increases.
Secondly, when the energy region of the 4d giant resonance is entered, 
the cross section increases sharply
because of the inter-shell coupling with the $4d$-$\epsilon f$ channel.

Fig. 5. shows the total photoionization cross section, which is equal to the sum of the 
$4d-\epsilon f$, $4d- \epsilon p$, $5s-\epsilon p$, and
$5p-\epsilon d, \epsilon s$ cross sections.  In Fig. 5 the solid curve represents our 
average of 
three terms, $4d^{10} 5s^2 5p^4 (^3P,\ ^1D,\ ^1S)$. The dashed curve is from reference [10] 
using the time-dependent local-density spin approximation with an optimized effective potential
and self-interaction correction (TDLSDA/OEP-SIC).
The dotted curve is the experimental data [14], and the dash-dot curve 
demonstrates the previous RPAE 
calculation [13].  
Our result shows a maximum of 23.12 MB at 90.24 eV, which is 
in excellent agreement with the recent experimental 
result, 23(3) MB at 90 eV [14].

In conclusion, we used our recently deveploped RPAE method to study the 
photoionization cross sections of the I$^+$ ion
with inclusion of all the inter-shell coupling among the $4d$, $5s$, and $5p$ subshells. 
We obtained, for the first time the 
photoionization cross sections for each term of the I$^+$ ground state.  
Our maximum cross section for the I$^+$ $4d$ giant resonance,  23.12 MB at the 
photon energy of 90.24 eV is in excellent 
agreement 
with the experimental data,
23(3)MB at 90 eV.

\bigskip
\leftline{\bf Acknowledgments}
\bigskip

This work was supported by the U.S. DOE, Division of Chemical Sciences,
Geosciences and Biosciences, Office of Basic Energy Sciences, OER. 
\bigskip

\leftline{\bf References}
\begin{enumerate}
\item M. Sano {\it et al}, 
J. Phys. B {\bf 29}, 5305 (1996).
\item E. D. Emmons {\it et al},
Phys. Rev. A {\bf 71}, 042704 (2005)  
\item T. Koizumi {\it et al}, 
Physica Scripta {\bf T73}, 131 (1997).       
\item N. Watanabe {\it et al},  
J. Phys. B:At. Mol. Opt. Phys. {\bf 31}, 4137 (1998).
\item Y. Itoh {\it et al},
J. Phys. B:At. Mol. Opt. Phys. {\bf 34}, 3493 (2001).
\item P. Andersen, T. Andersen, F. Folkmann, V. K. Ivanov, H. Kjeldsen and J. B. West, 
J. Phys. B:At. Mol. Opt. Phys. {\bf 34}, 2009 (2001). 
\item A. Aguilar {\it et al},
Phys. Rev. A {\bf 73}, 032717 (2006).
\item Zhifan Chen and A. Z. Msezane,
J. Phys. B: At. Mol. Opt. Phys. {\bf 39}, 4355 (2006).
\item M. Ya. Amusia, N. A. Cherepkov, L. V. Chernysheva and S. T. Manson,
J. Phys. B:At.Mol.Opt.Phys. {\bf 33}, L37-32 (2000)
\item A. T. Domondon and X. M. Tong, 
Phys. Rev. A {\bf 65}, 032718 (2002).
\item G. O'Sullivan, C. McGuinness, J. T. Costello, E. T. Kennedy, and B. Weinmann, 
Phys. Rev. A {\bf 53}, 3211 (1996).  
\item M. Ya. Amusia, N. A. Cherepkov, L. V. Chernysheva, and S. T. Manson,
Phys. Rev. A {\bf 61}, 020701(R) (2000)
\item M. Ya. Amusia, L. V. Chernysheva, V. K. Ivanov, and S. T. Manson,
Phys. Rev. A {\bf 65}, 032714 (2002). 
\item H. Kjeldsen, P. Andersen, F. Folkmann, H. Knudsen, 
B. Kristensen, J. B. West, and T. Andersen,
Phys. Rev. A {\bf 62}, 020702 (2000). 
\item L. Nahon, A. Svensson, and P. Morin,
Phys. Rev. A {\bf 43}, 2328 (1991). 
\item Zhifan Chen and A. Z. Msezane,
Phys. Rev. A, {\bf 77}, 042703 (2008). 
\end{enumerate}

\leftline{\bf Figure Captions}
Fig. 1. Photoionization cross section for the I$^+$ $4d-\epsilon f$ for the ground state terms of  
$(^3P,\ ^1D,\ ^1S)$. Dotted, dashed and dash-dot curves represent, respectively the results for   
$^3P$, $^1D$ and $^1S$.  

Fig. 2. Curves have the same meaning as those in Fig. 1, except that they demonstrate the I$^+$ 
$4d-\epsilon p$ photoionization cross section.
  
Fig. 3. Curves have the same meaning as those in Fig. 1, except that they demonstrate the I$^+$ 
$5p$ photoionization cross section.    

Fig. 4. Curves have the same meaning as those in Fig. 1, except that they demonstrate the I$^+$ 
$5s$ photoionization cross section.   

Fig. 5. Total cross sections for the I$^+$ photoionization in the energy region of $4d -\epsilon f$ 
giant resonance versus photon energy (eV). Solid curve is from our average 
cross sections for the terms 
$(^3P),\ (^1D)$, and $(^1S)$, dotted curve represents the experimental data, dashed curve shows the
calculated result from reference [10], and the dash-dot curve is from the 
calculation of Amusia et al [13].  
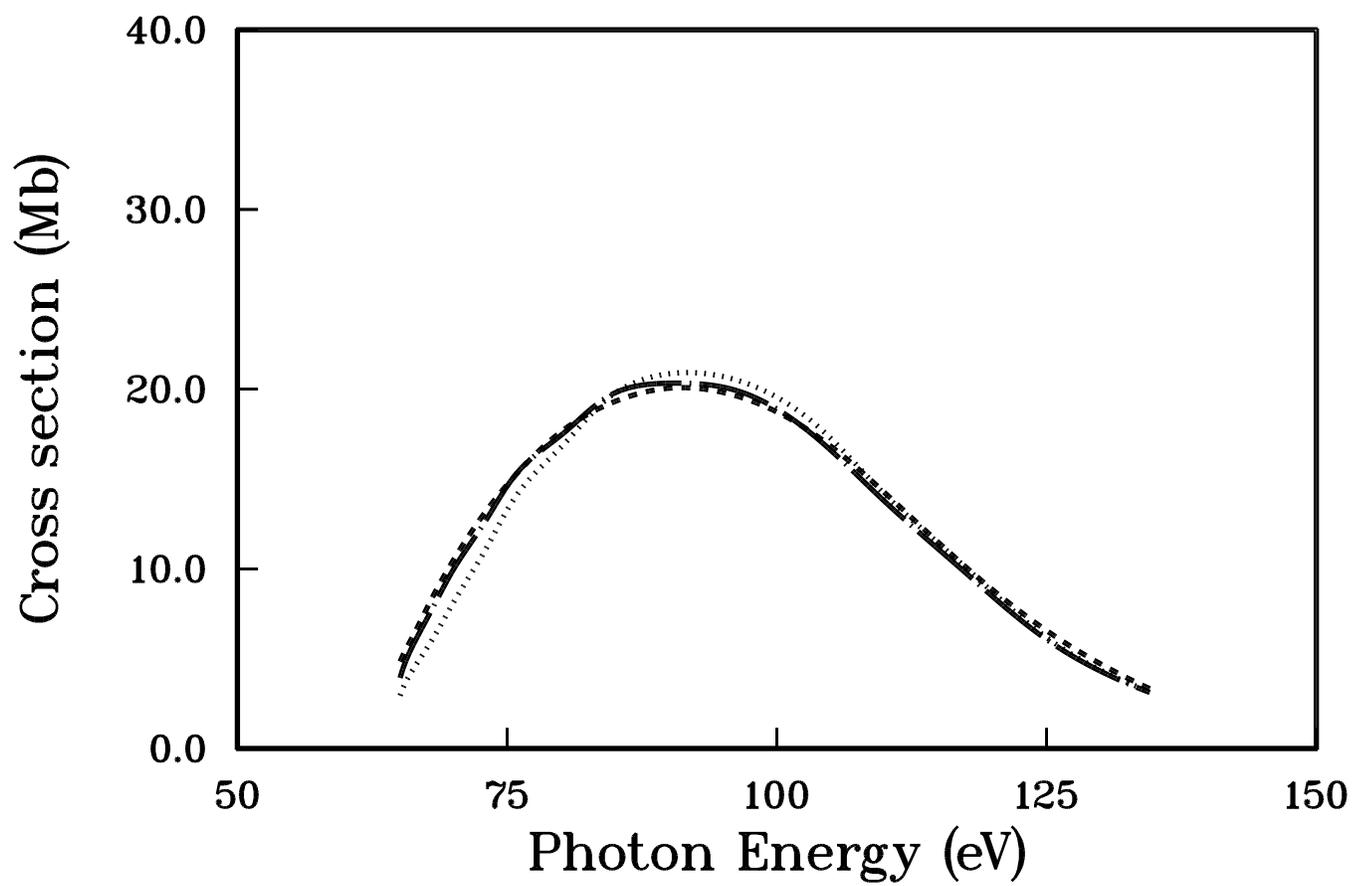
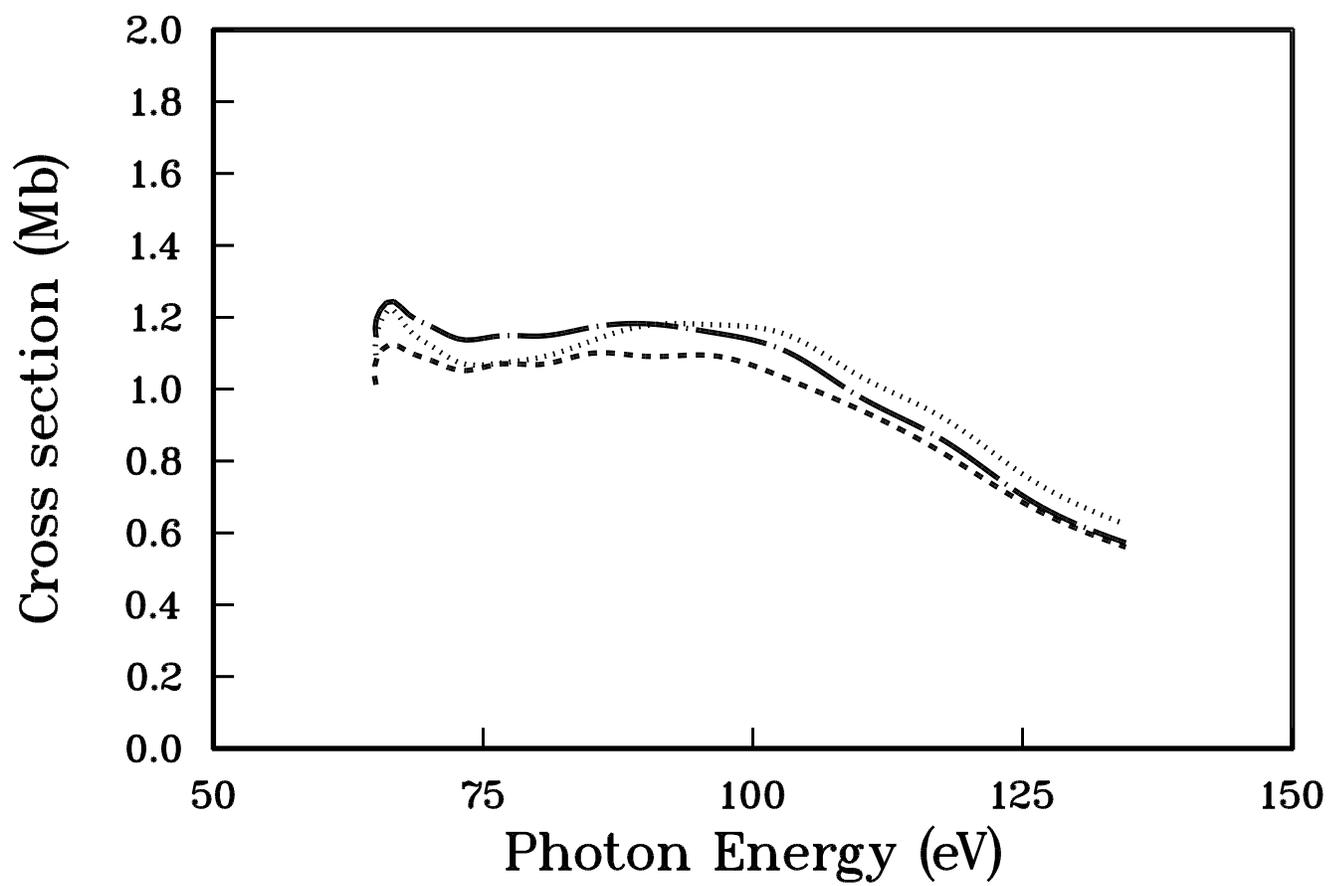
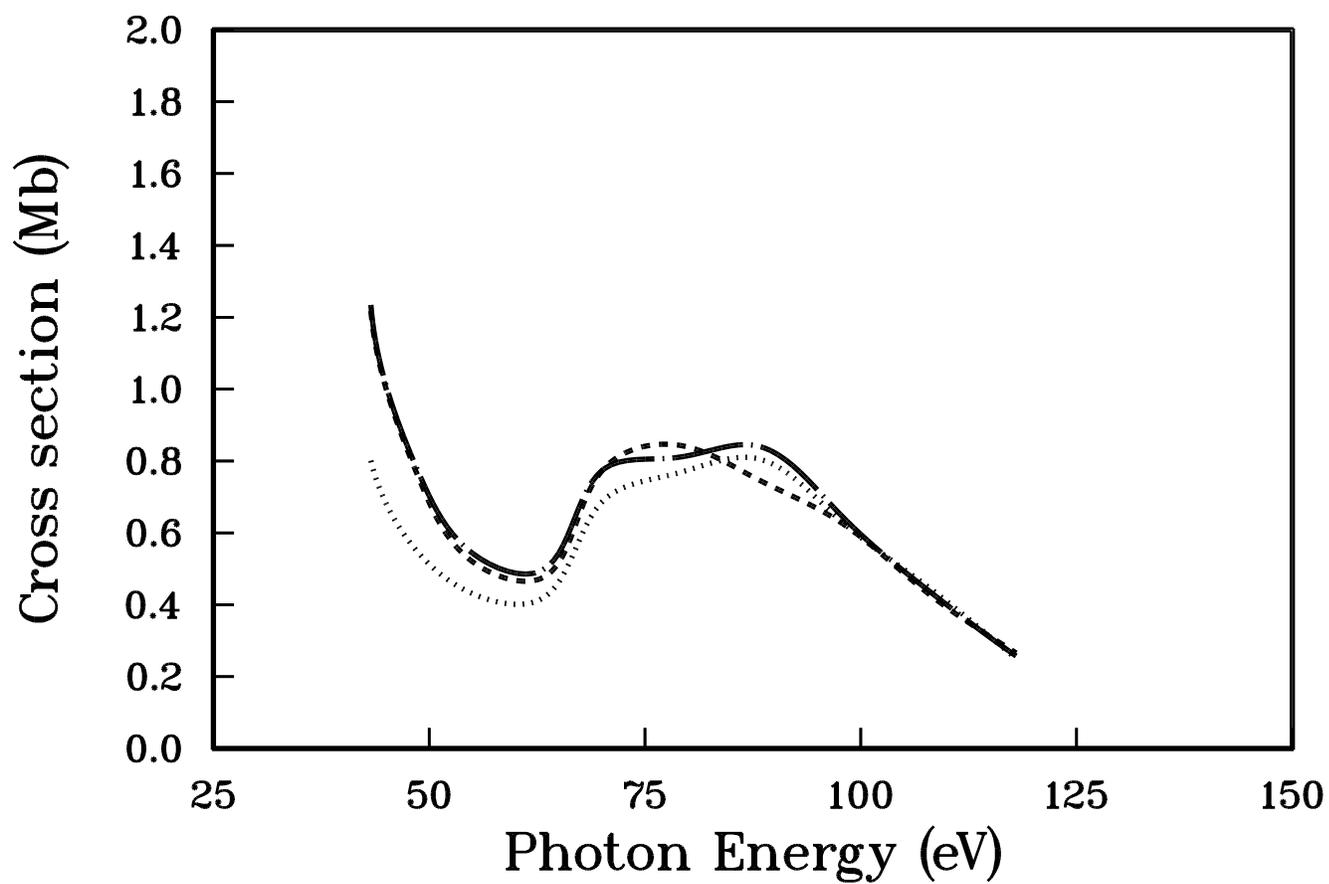
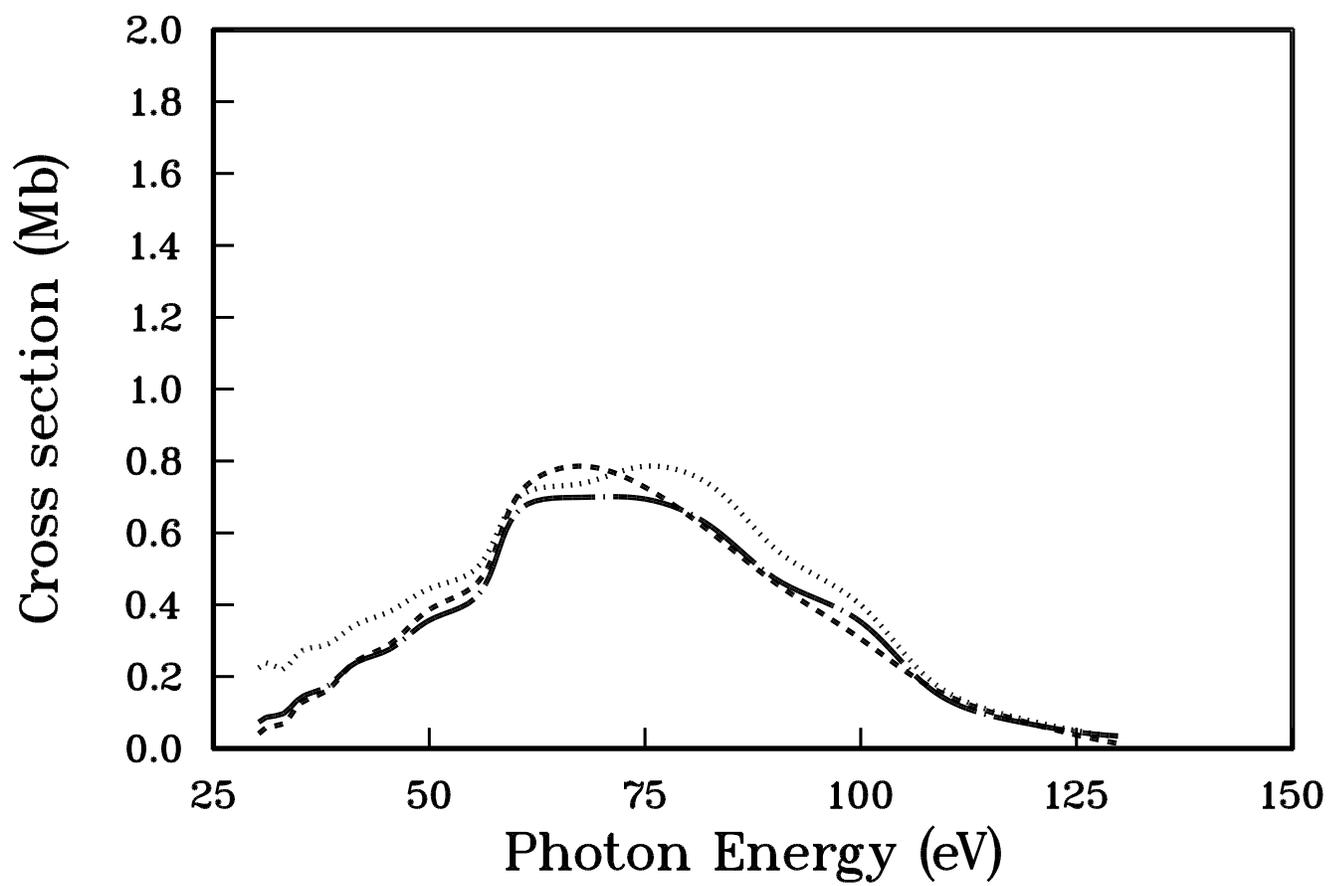
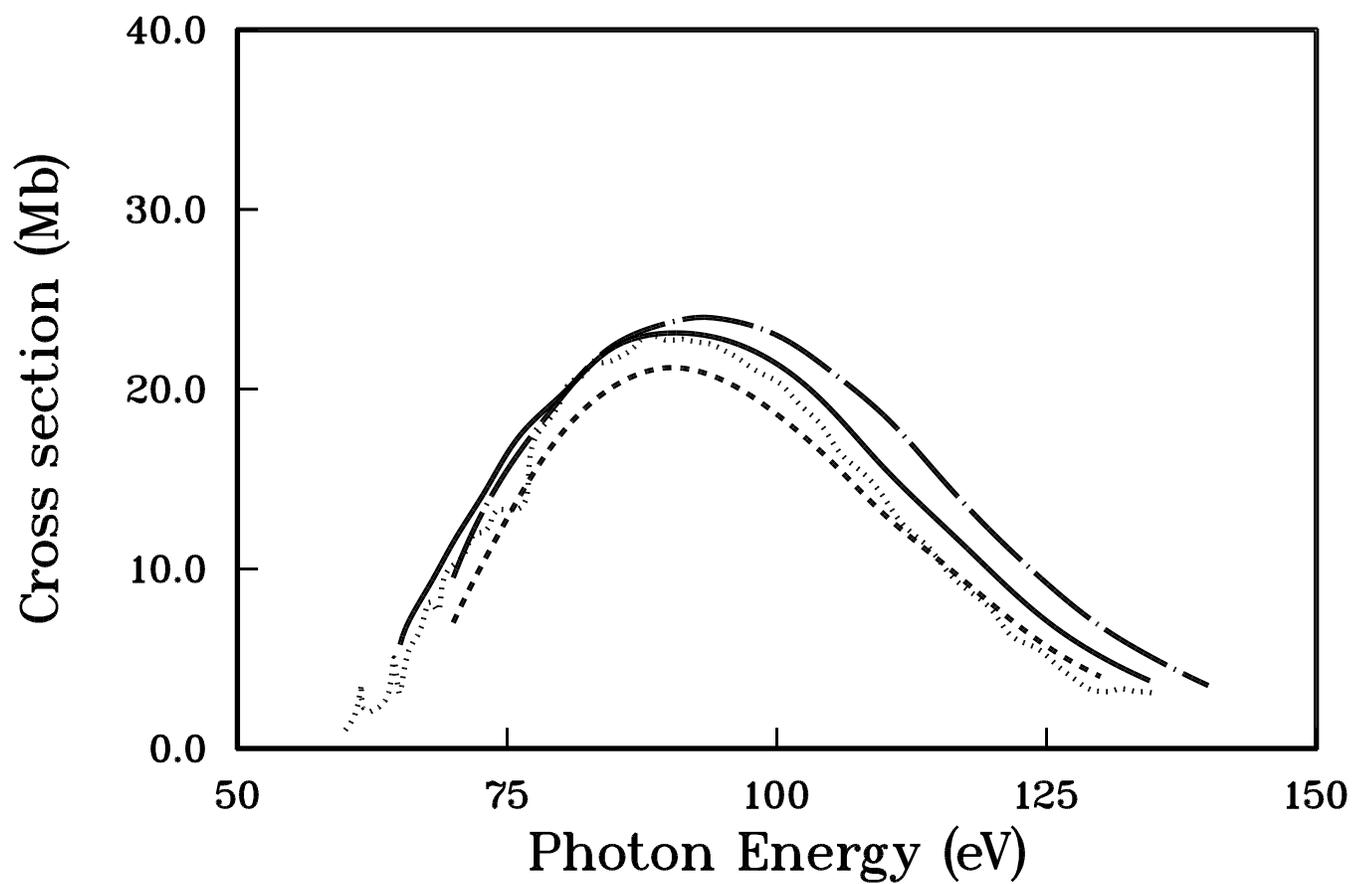
\end{document}